\documentclass{ametsocV6.1}

\title{Probabilistic Emulation of the Community Radiative Transfer Model Using Machine Learning\footnote{This Work has been submitted to Artificial Intelligence for the Earth Systems. Copyright in this Work may be transferred without further notice. This Work has not yet been peer-reviewed and is provided by the contributing Authors as a means to ensure timely dissemination of scholarly and technical Work on a noncommercial basis. Copyright and all rights therein are maintained by the Authors or by other copyright owners. It is understood that all persons copying this information will adhere to the terms and constraints invoked by each Author's copyright. This Work may not be reposted without explicit permission of the copyright owner.}}

\authors{Lucas Howard,\aff{a}\correspondingauthor{Lucas Howard, Lucas.Howard@Colorado.edu}
Aneesh C. Subramanian,\aff{a}
Gregory Thompson,\aff{b}
Benjamin Johnson,\aff{b}
Thomas Auligne,\aff{b}
}

\affiliation{\aff{a}{Department of Atmospheric and Oceanic Science, University of Colorado, Boulder}\\
\aff{b}{University Center for Atmospheric Research, Joint Center for Satellite Data Assimilation}\\
}

\abstract{The continuous improvement in weather forecast skill over the past several decades is largely due to the increasing quantity of available satellite observations and their assimilation into operational forecast systems. Assimilating these observations requires observation operators in the form of radiative transfer models. Significant efforts have been dedicated to enhancing the computational efficiency of these models. Computational cost remains a bottleneck, and a large fraction of available data goes unused for assimilation. To address this, we used machine learning to build an efficient neural network based probabilistic emulator of the Community Radiative Transfer Model (CRTM), applied to the GOES Advanced Baseline Imager. The trained NN emulator predicts brightness temperatures output by CRTM and the corresponding error with respect to CRTM. RMSE of the predicted brightness temperature is 0.3 K averaged across all channels. For clear sky conditions, the RMSE is less than 0.1 K for 9 out of 10 infrared channels. The error predictions are generally reliable across a wide range of conditions. Explainable AI methods demonstrate that the trained emulator reproduces the relevant physics, increasing confidence that the model will perform well when presented with new data.}

\begin{document}

\maketitle

%
%
%
%
%
%

%








\section{Introduction}
Operational weather forecast models provide key and actionable information to a wide range of stakeholders \citep{uccellini_evolving_2019} and the quality of these forecasts is highly sensitive to initial condition errors \citep{bauer_quiet_2015}. Observations from around the globe, including in-situ and remotely sensed data, are assimilated periodically to reset model initial conditions and correct forecast drift \citep{bannister_review_2017,edwards_regional_2015}. Data assimilation (DA) methods are used to perform this task.

The quality of initial conditions produced by DA is constrained by both the quality and quantity of assimilated observations. The proliferation of high-resolution satellite data has created both opportunities and challenges. Satellite radiance observations from various geostationary and low-earth-orbit platforms comprise most observations assimilated in operational NWP systems. These data can be assimilated as retrievals, with vertical atmospheric profiles inferred before being presented to the DA system, or as raw radiance observations. The latter method is most often used to simplify the representation of observation errors \citep{thepaut_satellite_2003}. Recently, advances have been made in the assimilation of all-sky radiance observations, where previously, primarily only clear-sky observations could be used in operational DA systems \citep{geer_all-sky_2018}.

Over the past several decades, advances in utilizing satellite radiance observations in DA systems have generated notable improvements in NWP forecast skill \citep{eyre_assimilation_2022,buehner_new_2018}. However, for some sensors, less than 1\% of available observations are assimilated \citep{johnson_community_2023} and generally, the vast majority of these observations are not used \citep{geer_all-sky_2018} for this purpose. The computational cost of the observation operator, used to map between model forecast variables and direct observations, is a significant bottleneck preventing the exploitation of a higher volume of available data \citep{johnson_community_2023}. With the expected availability of increasing numbers of hyperspectral instruments in the coming years, each with orders of magnitude more sensing channels than the current generation of sensors, the computational challenge limiting the exploitation of satellite observations will become more acute.

The Community Radiative Transfer Model (CRTM) is an example of a fast radiative transfer model and is widely used as an observation operator in DA frameworks. CRTM is developed and maintained by the Joint Center for Satelite Data Assimilation (JCSDA). It is a 1-D radiative transfer model designed to be fast and is used in DA to provide the expected radiance observed by various instruments, assuming the model forecast is correct \citep{johnson_community_2023}. Despite the advances made to speed up radiance computations in CRTM and other projects, the observation operator remains an impediment to assimilating increasing fractions of satellite data

Machine learning (ML) provides a potential path forward for this problem. ML techniques such as neural networks (NN) have been used increasingly in earth system science and adjacent applications, including in attempts to improve DA \citep{abarbanel_machine_2018,bonavita_machine_2021,penny_integrating_2022,sonnewald_bridging_2021,gettelman_future_2022}. This includes attempts to create fast observation operators \citep{geer_learning_2021,liang_machine_2023,liang_deep-learning-based_2022,stegmann_deep_2022,liang_applying_2020}. In particular, \citet{liang_applying_2020} and \citet{liang_deep-learning-based_2022} successfully trained NN models to emulate CRTM for two different instruments.

The NN-based emulators of CRTM cited above are accurate but only applicable to particular instruments. Training new ML models to allow for the assimilation of observations from other instruments is essential for fully realizing the benefits of this approach. Additionally, the NNs developed by \citet{liang_applying_2020} and \citet{liang_deep-learning-based_2022} are deterministic -- they produce a point estimate of the brightness temperature (or radiance) expected to be output by CRTM for a given set of input data. However, ML models can be trained to generate probabilistic outputs \citep{chapman_probabilistic_2022,howard_machine_2024,barnes_sinh-arcsinh-normal_2023}, which is valuable in a DA setting where observation operator errors must often be explicitly specified. Additionally, an ML emulator able to predict its error reliably could be used exclusively to assimilate observations where predictions are precise and expected errors are acceptably small. Observations could be discarded for other conditions, or existing observation operators such as CRTM used instead.

In this study, we develop an NN emulator of CRTM for a platform for which an ML operator does not yet exist and train it to produce probabilistic predictions. We target the Advanced Baseline Imager (ABI), a geostationary infrared instrument with spatial resolution ~2 km for infrared channels and temporal resolution of 5-15 minutes \citep{schmit_closer_2017}. Crucially, in NOAA's operational forecast system, only 0.02\% of ABI observations are assimilated largely due to the constraints imposed by CRTM's computational efficiency. We seek to show that a probabilistic NN can emulate CRTM for the ABI and produce reliable and valuable error estimates. Previous works that have built ML radiative transfer emulators have generated deterministic predictions, and the operational workflow suggested here with simultaneous use of the emulator and CRTM would therefore not be possible.

Following the introduction, the rest of this paper is organized as follows: section 2 describes the methods and data used, including CRTM and generation of training data, NN architecture, training and tuning of the NN, and explainable AI techniques used to analyze the trained model. Section 3 presents the training results and the NN's performance. Section 4 summarizes the results and discusses implications for future work, with section 5 concluding.
\section{Data and Methods}\label{methods}
\subsection{CRTM}
CRTM is a 1-D radiative transfer model capable of simulating scattering and absorption of a wide variety of atmospheric constituents and hydrometeors, and of predicting the radiance observed by many different sensors given specified atmospheric conditions. Designed for speed and use in DA systems, it uses lookup tables for calculating optical properties of aerosols and clouds and a regression model fit to line-by-line models to speed up transmittance calculations. JCSDA maintains it, and development is ongoing with additional capabilities expected to be included in new releases \citep{johnson_community_2023}. While CRTM is designed for computational efficiency, it and other models like it continue to be a bottleneck preventing the utilization of more available data. As a result, synergies between CRTM (capable of producing large volumes of training data offline) and machine learning are already envisioned for the future of observation operators within DA systems \citep{johnson_community_2023}.

\subsection{Data}\label{data}
To generate input data for CRTM and the NN, the Global Forecast System Finite Volume Cubed Sphere (GFS FV3) was used. ABI data was simultaneously subsampled to a uniform grid with 64 km spacing. Forecast data from the cube-earth grid was horizontally interpolated to the locations of the subsampled ABI data points, with both surface data and vertical profile data extracted from the forecast at these locations. CRTM was then used to generate predicted radiances for channels 7-16, converted to brightness temperatures via the Planck function. For simplicity, this work does not target channels 1-6 which cover visible wavelengths, and instead exclusively focuses on near-IR and IR channels.

30 days of simulated scans at 6-hour intervals were generated for both the GOES-16 and GOES-17 platforms. JEDI-Skylab was used to create the data, and the system does not generate simulated sensor data when an observation does not exist. As a result, while the actual observed data are not used for training and the non-existence of observation for a particular time interval is therefore not relevant to the ML training process, CRTM-simulated ABI observations are not available for every time step for both GOES-16 and GOES-17. In total, 151 simulated scans were generated for February 15, 2022-March 15, 2022.

\subsection{NN Architecture}
The NN takes identical input variables as CRTM, which for this application includes nine atmospheric variables at each of the 127 pressure levels of the GFS, 16 surface variables, and seven metadata variables. These are listed in Table \ref{t1}. Output variables are predicted brightness temperatures for channels 7-16 and predicted error standard deviations for each channel (with the error referring to disagreement with CRTM predictions). This is consistent with the approach taken by e.g. \cite{chapman_probabilistic_2022} and \cite{howard_machine_2024} in designing a NN that generates probabilistic predictions.

\begin{table}[t]
\begin{center}
\begin{tabular}{ccccrrcrc}
\hline\hline
Input Variable Name & Variable Type & Number of Variables \\
\hline
 Air Temperature & Atmospheric & 127 \\
 Water Vapor Mixing Ratio & Atmospheric & 127 \\
 Cloud Ice Mass &  Atmospheric & 127 \\
 Cloud Water Mass  &  Atmospheric & 127 \\
 Snow Mass &  Atmospheric & 127 \\
 Ice Particle Effective Radius & Atmospheric & 127 \\
 Water Particle Effective Radius & Atmospheric & 127 \\
 Snow Particle Effective Radius &  Atmospheric & 127 \\
 Ozone Mixing Ratio &  Atmospheric & 127 \\
 Ice Area Fraction & Surface & 1 \\
 Land Area Fraction & Surface & 1 \\
 Land Type & Surface & 1 \\
 Leaf Area Index & Surface & 1 \\
 Soil Temperature & Surface & 1 \\
 Snow Area Fraction & Surface & 1 \\
 Snow Thickness & Surface & 1 \\
 Soil Temperature & Surface & 1 \\
 Surface Temperature (Ice, Land, Sea, Snow) & Surface & 1 \\
 Wind Direction & Surface & 1 \\
 Wind Speed & Surface & 1 \\
 Vegetation Area Fraction & Surface & 1 \\
 Water Area Fraction & Surface & 1 \\
 Volumetric Water Ratio (soil) & Surface & 1 \\
 Sensor Scan Angle & Meta & 1 \\
 Sensor Zenith Angle & Meta & 1 \\
 Sensor View Angle & Meta & 1 \\
 Sensor Azimuth Angle & Meta & 1 \\
 Sensor Elevation Angle & Meta & 1 \\
 Solar Azimuth Angle & Meta & 1 \\
 Solar Zenith Angle & Meta & 1 \\
\hline
&Total Input Variables: & 1,166
\end{tabular}
\end{center}
\caption{List of input variables for CRTM needed to predict the radiance observed by the ABI.}\label{t1}
\end{table}

The network is comprised of 3 hidden fully connected layers with 512 nodes in each hidden layer. While other architectures have been shown to have superior accuracy for some radiative transfer problems such as bidirectional recurrent neural networks \citep{ukkonen_exploring_2022}, this performance comes at the cost of computational efficiency. For this application where computational efficiency is key, a simpler dense NN is an appropriate choice.

The hidden layer activation function is SWISH, defined as:

\begin{equation}
f(x) = x \cdot\text{sigmoid}(\beta x) = \frac{x}{1+e^{-\beta x}}
\end{equation}
where $\beta$ is a trainable parameter. The output layer activation for the predicted brightness temperatures is sigmoid and output targets are scaled to between 0 and 1 via:

\begin{equation}
x_{scaled} = \frac{x_{raw}-\beta_{min}}{\beta_{max}-\beta_{min}}
\end{equation}
where $\beta_{min}$ and $\beta_{max}$ are tunable parameters representing the minimum and maximum brightness temperatures that the network can predict. The final values used in the results presented here are 180 K and 355 K, respectively.

The output layer activation for the predicted standard deviations is:

\begin{equation}
    f(x)=\delta+\alpha\cdot\text{softmax}(x)
\end{equation}
with $\delta>0$ is a tunable parameter and $\alpha$ is trainable. $\alpha$ is a scaling parameter, and allowing it to vary during training resulted in faster convergence by shifting the center of the output activation distribution to the true center of the target distribution. $\delta$ is a small value defining the minimum standard deviation (maximum precision) that the network can predict. This improved the stability of the training, allowing larger initial learning rates to be used and resulting in faster convergence. The final value was set to 0.001 K.

\subsection{Training, Validation, and Tuning}
The dataset described in section \ref{methods}\ref{data} was split into training, validation, and test sets in an approximate ratio of 80/10/10, corresponding to 121/15/15 scans. The split was generated randomly. The validation training set was used during training to monitor the predictive performance of the NN model and prevent overfitting, as well as for hyperparameter tuning. Still, it was not used in computing the gradient for adjusting weights. The test dataset was held back until hyperparameter tuning and training were complete and used to evaluate the generalization performance of the model.

The Adam optimizer \citep{kingma_adam_2015} as implemented in Keras \citep{chollet_keras_2015} was used to train the model using a batch size of 32 and a maximum of 200 epochs. The gradient at each update step was clipped to a maximum value of 0.5. The learning rate was adaptively reduced during training; if the validation loss value does not decrease by 0.01 for 3 epochs in a row, the learning rate is reduced by a factor of 5 to a minimum learning rate of 1e-6. Early stopping was also employed, with training automatically terminating if the validation loss did not decrease by more than 1e-5 for 10 straight epochs. L2 weight regularization was also employed in hidden layers to improve training stability and reduce the risk of overfitting.

The loss function used is the continuous rank probability score (CRPS), a scoring function used for probabilistic predictions \citep{gneiting_strictly_2007}. The neural network predicts a normal distribution of the expected CRTM-predicted brightness temperature. CRPS values will be higher for imprecise predictions (where the spread is large) or inaccurate (where the difference between the predicted mean and target is large). NN weights that successfully minimize mean CRPS across the training dataset will, therefore, result in a probabilistic CRTM emulator that is both accurate and well characterizes its error with respect to the target dataset.

The optimal configuration of NN architecture, activation function(s), and training settings are generally not evident \emph{a priori} and tuning of these values is often necessary. Tuning of these hyperparameters can be automated using various optimization methods, but for this study, automated hyperparameter tuning was not considered necessary. Instead, hyperparameters were adjusted manually, and a set of hyperparameters was chosen based on the performance of the trained models in the validation training dataset. The hyperparameters adjusted in this process are listed in Table \ref{hyperparams} along with a description and the final value.

\begin{table}
    \begin{center}
    \begin{tabular}{ccc}
    \hline
    \hline
         Hyperparameter Name&  Description& Final Value\\
         \hline
         Network Depth&  Number of hidden layers in the NN& 3\\
         Nodes per Layer&  Number of hidden nodes in each hidden layer& 512\\
         $\lambda_{2}$&  L2 regularization factor& 1e-8\\
          nepochs& Maximum number of training epochs& 200\\
          batchsize&Number of samples evaluated per training batch & 32\\
          $L_{0}$& Initial training rate & 5e-4\\
         $\delta$&  Minimum predicted standard deviation in units of K& 0.001\\
 $\beta_{min}$& Output brigthness temperature normalization offset&180\\
 $\beta_{max}$& Output brightness temperature normalization scaling factor&355\\
    \end{tabular}
    \caption{List of NN hyperparameters, a description of the hyperparameter, and the final value chosen after manual tuning.}
    \label{hyperparams}
    \end{center}
\end{table}

\begin{figure}[t]
  \centering
  \noindent\includegraphics[width=25pc,angle=0]{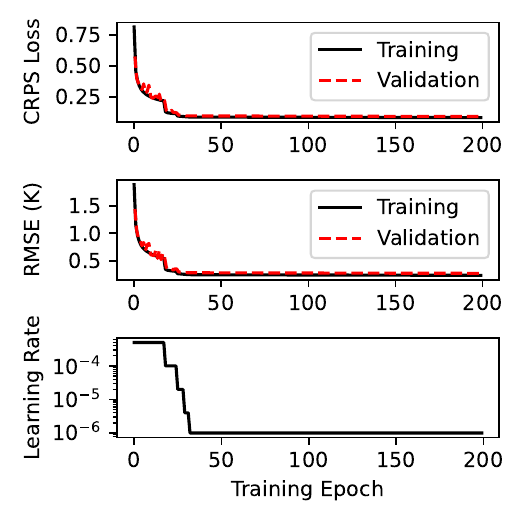}\\
  \caption{CRPS loss (top) and RMSE (middle) are shown as a function of training epoch. Results for both training data (in solid black) and validation data (in dashed red) are included. The learning rate is also shown (bottom)}
  \label{r1}
\end{figure}

The training process results are shown in Figure \ref{r1}. CRPS loss (top) and RMSE (middle) are included for both the training dataset as well as the validation dataset. Both metrics are averaged across all channels. Predictive performance on the validation dataset (not used in training) is monitored to prevent overfitting. Both training and testing performance are consistently flat following multiple decreases in the learning rate; this indicates that a local minimum has been reached and that the model is not overfit. Overall validation RMSE is less than ~0.3 K, which is the same order of magnitude error generated by \citet{liang_deep-learning-based_2022} and is considered acceptable at this relatively early stage of development.

\subsection{Explainable AI}
Using a NN model as an emulator of a physics-based operator may be much more computationally efficient once the model is trained, allowing for the exploitation of many more observations. This has obvious benefits for operational DA systems. One major drawback of this approach is that NN models are effectively black boxes which provide no obvious indication of how they generate predictions \citep{gevaert_explainable_2022}. Additional steps beyond evaluating accuracy on validation or test datasets may therefore be needed to provide confidence in the performance of the trained model may therefore be needed before deploying it in an operational setting.

Explainable AI (XAI) methods are tools that can make the behavior of black box models more transparent and provide insights into how the model generates its predictions \citep{linardatos_explainable_2021}. Shapely Additive Explanations (SHAP) is an XAI method that quantifies the impact of input variables on model outputs, assigning each input variable credit for a portion of the output deviation from a baseline expectation value. A formal definition and full details of the method can be found in \cite{lundberg_unified_2017}. In machine learning and DA applications, SHAP has previously been used to analyze a trained neural network \citep{howard_machine_2024}.

Here we apply SHAP to a randomly selected sample of 1000 clear-sky and 1000 cloudy pixels. The presence or absence of clouds is a large contributor to observed radiance, and splitting the dataset in this way aids in interpreting the SHAP results. Cloudy conditions were identified based on cloud water, cloud ice, or snow in the atmospheric column in the input data.

\section{Results}
\subsection{Predictive Performance and Benchmarking}

\begin{figure}[t]
  \centering
  \noindent\includegraphics[width=\linewidth,angle=0]{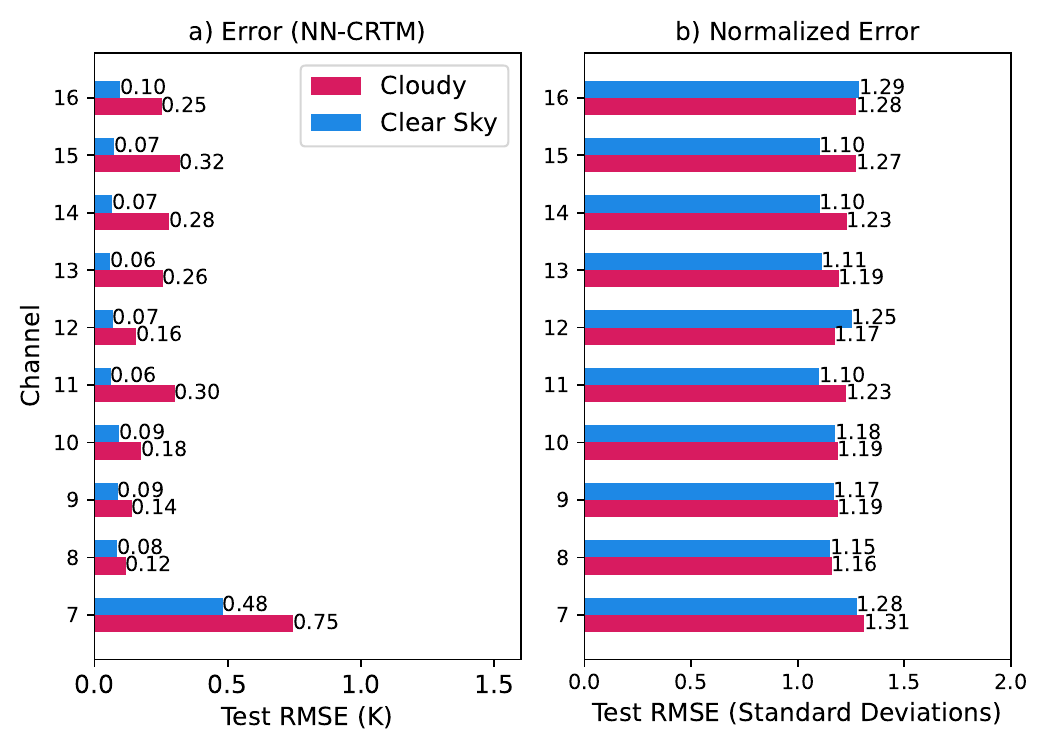}\\
  \caption{RMSE for all channels in both cloudy and clear-sky conditions are shown in panel a). RMSE of the normalized error is shown in panel b).}
  \label{r2}
\end{figure}

Figure \ref{r2} shows the RMSE on the test dataset by channel. The RMSE is shown on the left for both clear sky and cloudy conditions. Channel 7, which has a significantly reflected short-wave IR component, has a notably larger error than the other channels, which do not have a reflected component of their radiances. Cloudy conditions also tend to produce larger errors, which is consistent with the more complicated scattering dynamics involved.

The normalized RMSE by channel is also included on the left. To produce these values, errors were divided by the predicted standard deviation. If the actual errors are well characterized by the NN, the normalized errors will be a unit normal distribution, and the RMSE will be 1. The normalized RMSE is close to but larger than 1 for all channels and conditions, indicating that the NN slightly underpredicts the true errors.

\clearpage
\begin{figure}[t]
  \centering
  \noindent\includegraphics[width=4in]{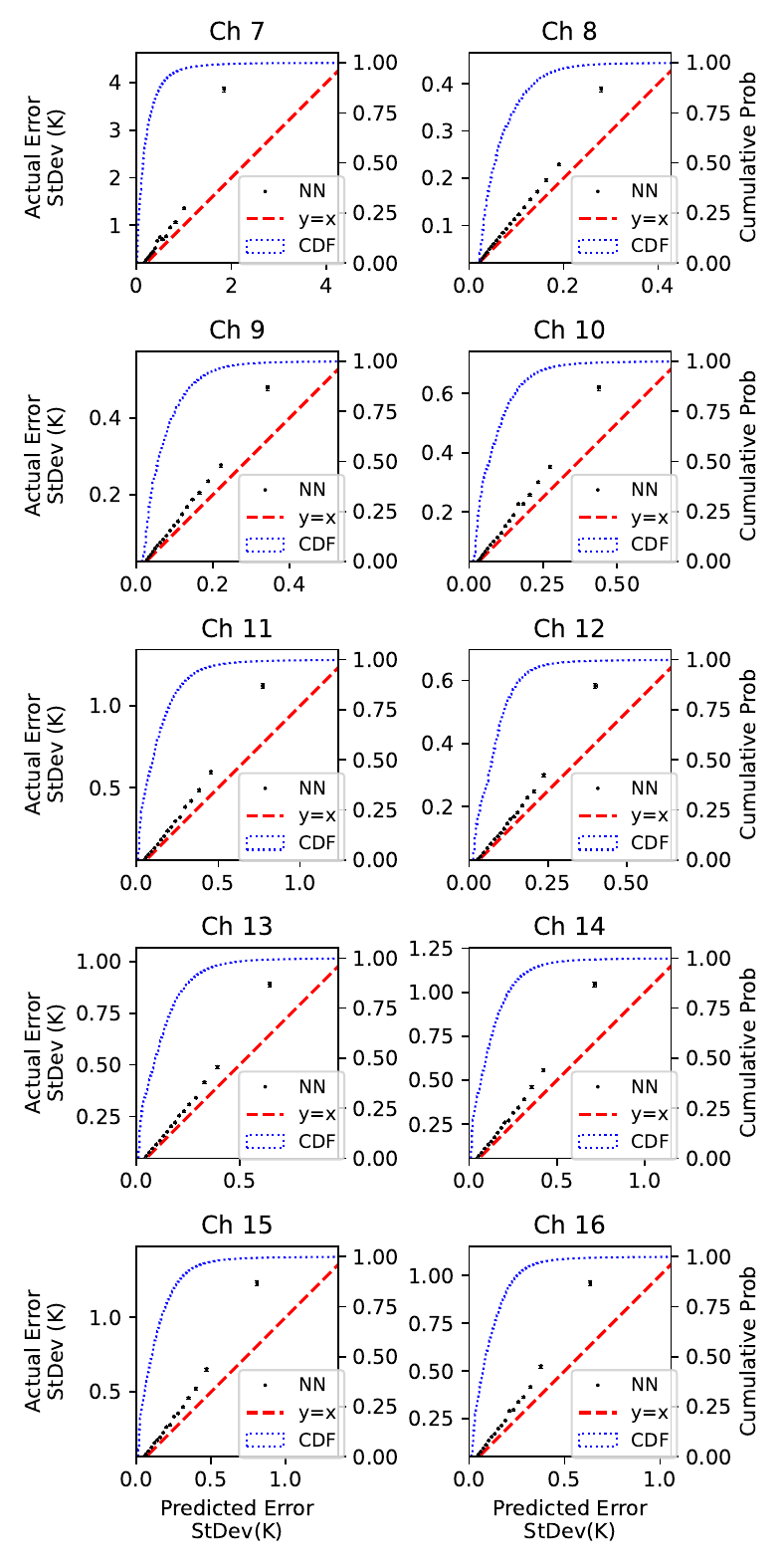}\\
  \caption{Calibration curves of the NN predictions are shown for all channels with 95\% error bars (black). The predicted error standard deviation is on the x-axis, with the actual error standard deviation on the y-axis. The y=x line is also shown (dashed red) along with the cumulative probability distribution of the predicted error (dotted blue).}
  \label{r3}
\end{figure}
\clearpage

Probabilistic predictions can be evaluated based on how well they characterize events (i.e. if the predicted probabilities and true probabilities are aligned) and resolution (the range of predicted probabilities generated). A model that predicts the climatological mean and standard deviation for all cases will characterize its errors well but with poor resolution. Figure \ref{r2}b) provides insights about the errors' overall distribution compared to the NN's expectations. Figure \ref{r3} gives information about the resolution of the predictions as calibration curves.

The predicted error standard deviations (x-axis) are binned and plotted against the actual error standard deviation for each channel. 95\% confidence intervals in both directions and an idealized curve (y=x) as a red dashed line are included. For smaller errors, the predictions are very well calibrated, while for larger errors the NN tends to underestimate the true error systematically. However, the general trend is monotonic -- larger predicted error standard deviations tend to have larger errors. Additionally, compared to the RMSE's in Figure \ref{r2}a) the predictions are well calibrated for a wide range of true errors and only become notably separated from the ideal curve for values roughly an order of magnitude greater than the RMSE. This is true for channel 7 as well, despite the apparent dramatic underestimation evident at larger errors.

\clearpage
\begin{figure}[t]
  \centering
  \noindent\includegraphics[width=6in,angle=0]{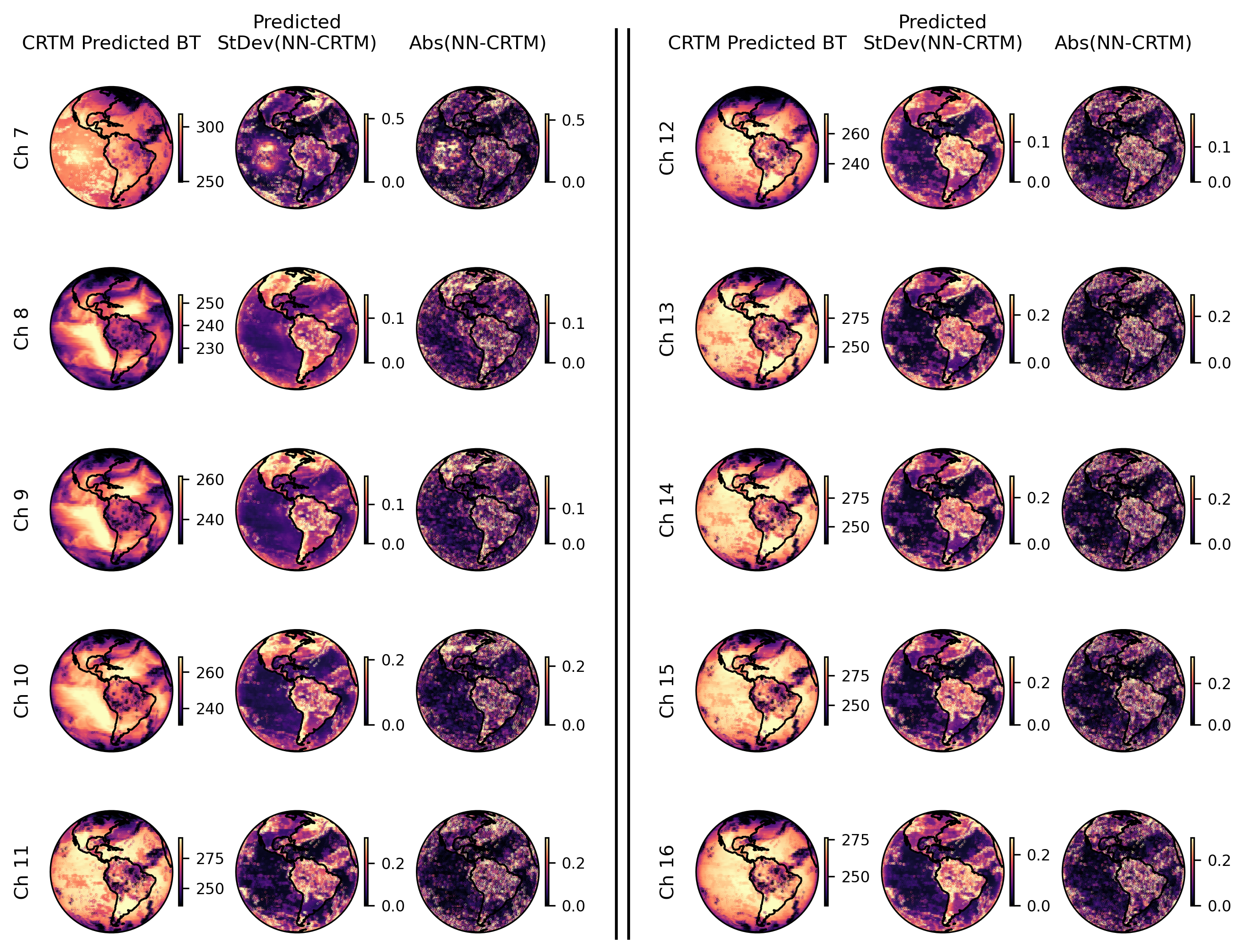}\\
  \caption{For one of the test scans, maps of the CRTM-predicted brightness temperature, predicted error standard deviation, and actual absolute error (left to right) are shown for each channel.}
  \label{r4}
\end{figure}
\clearpage

One final way to visualize the predictive performance of the NN emulator is with a map of its predictions compared to CRTM for a single scan. This is shown for one of the images from the test dataset in Figure \ref{r4}. The CRTM-predicted brightness temperature (the value that the NN is trying to match) is shown along with the predicted error standard deviation and actual absolute error. As the NN predicts error distributions, rather than a specific value, we do not expect perfect alignment between the second column (predicted error standard deviation) and the third column (actual absolute error). The third should be noisier, and this is what is observed. Spatial patterns line up well though, will regions with large (or small) predicted errors tending to have actual errors that are relatively larger (or smaller).

\begin{figure}[t]
  \centering
  \noindent\includegraphics[width=3in,angle=0]{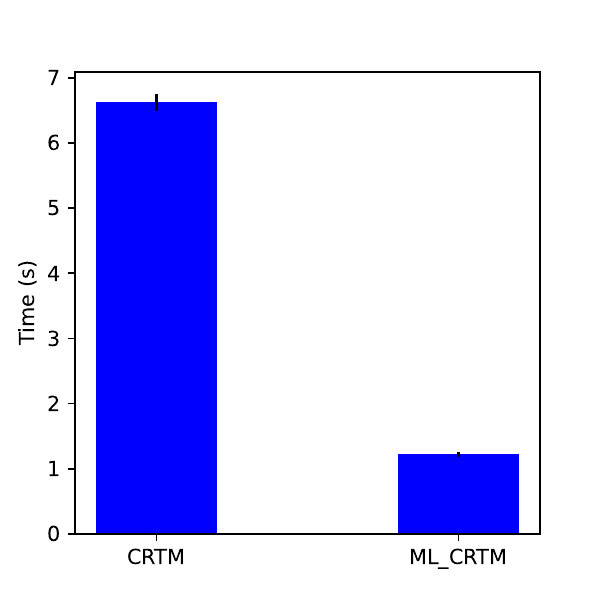}
  \caption{The mean execution time across 100 runs performed on a single scan is shown for CRTM (left) and the ML model (right). 95\% confidence intervals are included for both models.}
  \label{r5}
\end{figure}

Additionally, as computational efficiency is one of the main motivations for this work, both CRTM and the trained emulator were run on a single scan 100 times on a single CPU.  GPU acceleration or other parallelism was not used. Although this is not representative of an operational setting, using GPU acceleration (for the ML model) or parallelization with MPI (for CRTM) would make it difficult to compare the relative efficiency of the two models. The results are shown in figure \ref{r5}. With limited effort used to speed-optimize the code, the ML model is faster than CRTM by approximately a factor of 5. CRTM is also substantially slower for cloudy conditions, meaning that the speed-up provided by the ML model would be even higher for these data points.

\begin{figure}[t]
  \centering
  \noindent\includegraphics[width=6in,angle=0]{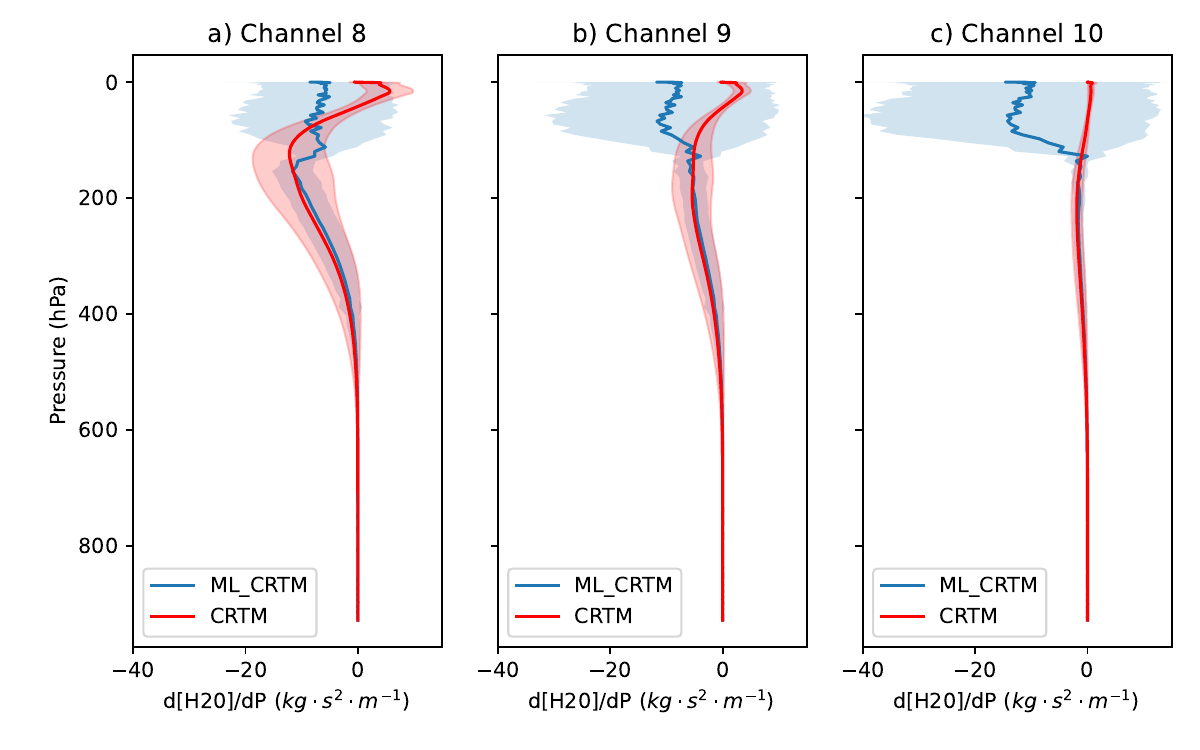}
  \caption{Mean Jacobians plus or minus a standard deviation for the upper-level (channel 8, panel b), mid-level (channel 9, panel c) and lower-level (channel 10, panel c).}
  \label{r6}
\end{figure}

Finally, we compare the Jacobians of the water vapor channels produced by the ML model to those of CRTM. Stable, accurate Jacobians are critical if the ML emulator is to be used in an operational DA setting as many DA methods require adjoint versions of the observation operators. We computed the derivatives of both CRTM and the ML emulator for the water vapor channels with respect to water vapor mixing ratio at each pressure level for a single scan. Following \cite{liang_deep-learning-based_2022} we plot the mean value along with error bars representing one standard deviation. These results are shown in Figure \ref{r6}. For lower altitudes, the mean and spread match relatively well. Above 200 hPA the ML model Jacobian diverges significantly from CRTM. This is likely due to the lack of variability in stratospheric water vapor variability. 

\subsection{Explainable AI}
We use explainable AI in this work to confirm that the NN is learning the relevant physics and is consistent with our expectations to increase confidence in using it on other datasets If, for example, highly correlated channels have widely divergent SHAP values, this would be an indication that the network is not generating predictions based on factors that we are confident would be reproduced for an arbitrary set of new input data. Conversely, if those channels have similar SHAP values, this would indicate that the network is using similar physical features to generate predictions for the correlated channels and would tend to increase confidence in the future performance of the trained model.

\begin{figure}[t]
  \centering
  \noindent\includegraphics[width=22pc,angle=0]{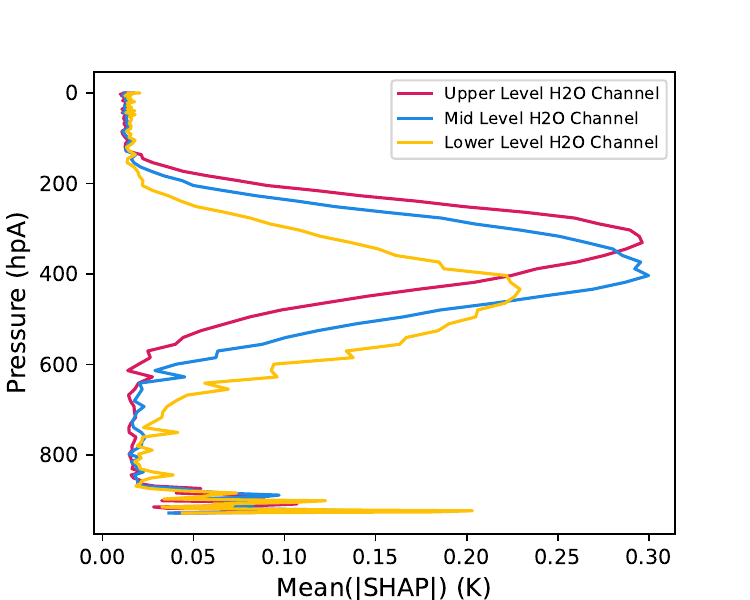}\\
  \caption{For channels 8, 9, and 10 (the upper, mid-level, and lower-level water vapor channels) the mean absolute SHAP value is shown as a function of pressure level.}
  \label{r7}
\end{figure}

For each combination of input and output variable, there could be a corresponding SHAP analysis. We focus first on the impact of atmospheric water vapor channels 8, 9  and 10 -- also known as the upper-level, mid-level, and lower-level water vapor channels. For these channels, there is a clear qualitative expectation of the results if the NN has learned the proper physics. Vertical profiles for all the mean absolute SHAP value at all pressure levels are shown in Figure \ref{r5} for clear sky conditions. The results in this plot clearly show the peaks where they would be expected for the three channels, highest in the atmosphere for the upper-level channel and lowest for the lower-level channel. This figure also suggests that the divergence in the Jacobians above 200 hPA is unlikely to be problematic as water vapor at this altitudes has nearly zero impact on simulated brightness temperatures.

\begin{figure}[t]
  \centering
  \noindent\includegraphics[width=30pc,angle=0]{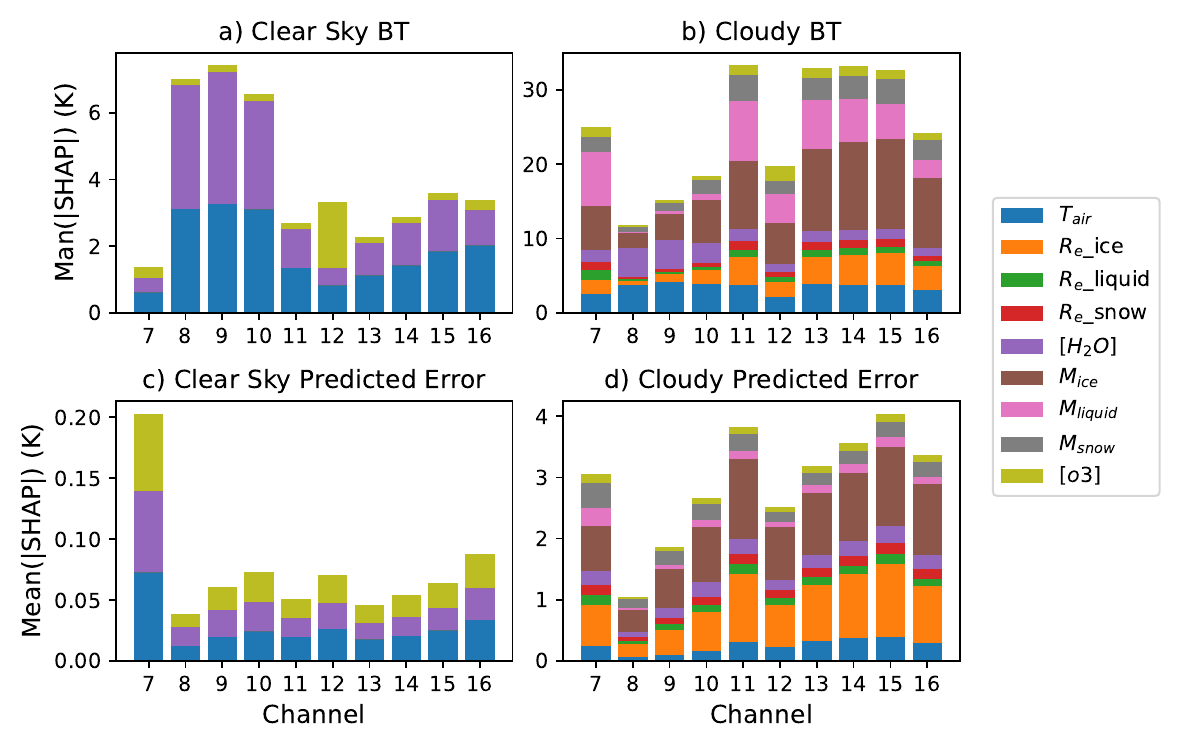}\\
  \caption{The mean absolute vertically integrated SHAP values for each atmospheric variable are shown as stacked bar plots for each channel. Panel a) shows the impact on predicted brightness temperature for clear sky conditions. Panel b) shows the impact on brightness temperature for cloudy conditions. Panel c) shows the impact on the predicted error for clear sky conditions and d) on the predicted error for cloudy conditions.}
  \label{r8}
\end{figure}

We can also look at the SHAP values for all atmospheric variables vertically integrated overall pressure levels. This provides information about the impact each variable overall rather than at specific pressure levels. These results are shown in Figure \ref{r6}, broken down into clear sky and cloudy conditions. Each variable's impacts on the predicted brightness temperature and error standard deviation are included. The impact of water vapor on predicted brightness temperature is, as expected, the largest in both clear sky and cloudy conditions for the water vapor channels (8, 9, and 10). Ozone greatly impacts channel 12 brightness temperature (the ozone channel), again consistent with expectations. Various cloud variables contribute heavily to the error in all channels.

\begin{figure}[t]
  \centering
  \noindent\includegraphics[width=30pc,angle=0]{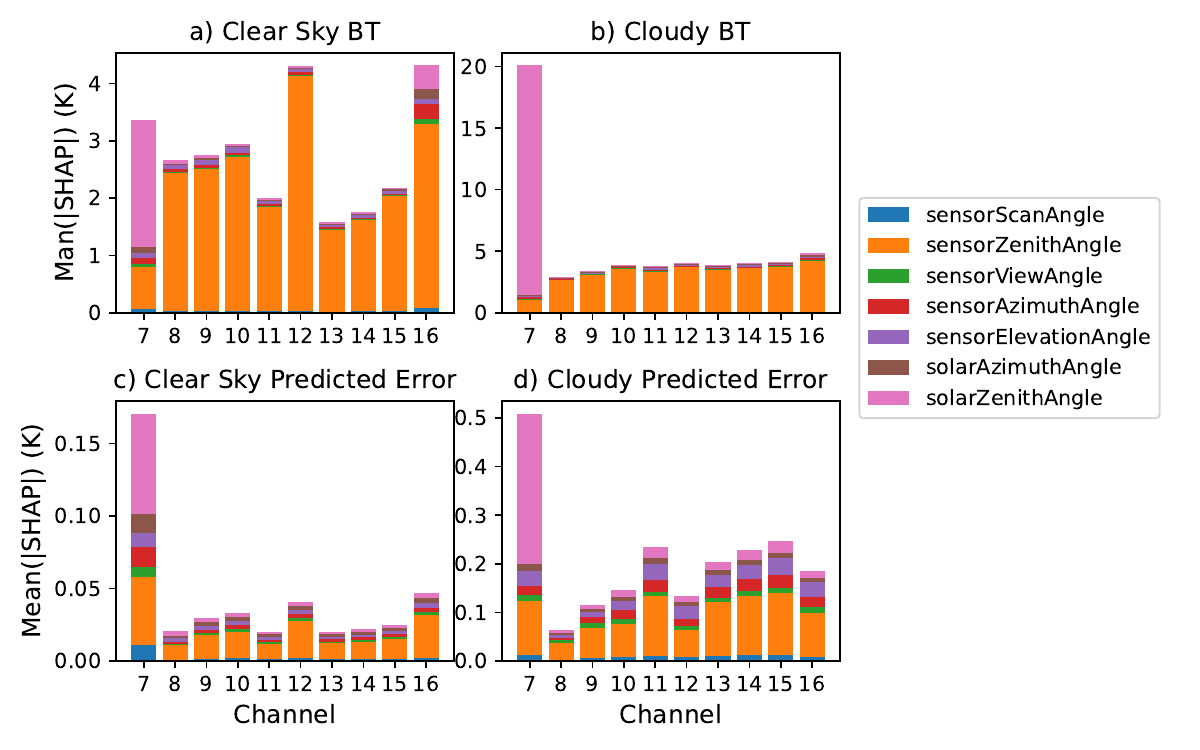}\\
  \caption{The mean absolute SHAP values for each of the meta variables are shown as stacked bar plots for each channel. Panel a) shows the impact on predicted brightness temperature for clear sky conditions. Panel b) shows the impact on brightness temperature for cloudy conditions. Panel c) shows the impact on the predicted error for clear sky conditions and d) the impact on predicted error for cloudy conditions.}
  \label{r9}
\end{figure}

We also examine the impact of the meta variables, which include sensor and solar angles. As in Figure \ref{r6}, Figure \ref{r8} has results broken down by clear-sky and cloudy and includes the impact of each variable on predicted brightness temperature and error standard deviation. The results in this figure that stand out are the SHAP values for channel 7. Solar zenith angle has a very large impact on both predicted brightness temperature and predicted error in both clear-sky and cloudy conditions. As was discussed earlier, channel 7 includes significant reflected  short wave IR, and the solar zenith angle is a large factor in the quantity of reflected solar radiation observed by the sensor. This result is again consistent with expectations of the behavior the the radiative transfer physics. It increases our confidence that the NN has learned the true physics rather than memorizing the dataset.

\section{Discussion}
The intended use case of the probabilistic nature of the NN is to facilitate integration into operational DA systems. The probabilistic prediction is of little use. However, if the predictions are not sufficiently accurate for there to be a reasonable expectation of improved DA performance using the NN operator. The first question that must be answered is: does the trained NN produce sufficiently accurate predictions to be used as an observation operator in a DA system? The error predictions in turn must be analyzed with respect to the use cases of the probabilistic component of the NN. Two potential applications are envisioned: first, to use a threshold predicted error above which NN results will not be used, and second, as input into the observation error term. Figures \ref{r2} and \ref{r3} can be examined to answer these questions.

First considering only the deterministic prediction, the RMSE across all channels is ~0.3 K, slightly larger but comparable to the results generated by \cite{liang_deep-learning-based_2022} in developing a NN emulator of CRTM for a different instrument. Much of this error is driven by channel 7 (Figure \ref{r2}a), and in particular, the clear sky performance for the other channels is consistently $<$0.1 K. This level of accuracy is judged to be sufficient to support conducting future DA experiments using the NN emulator.

Figure \ref{r2}b) provides evidence that the overall error distribution is well characterized across the test dataset. If the true distribution of errors is aligned with the predicted standard deviations, the RMSE will be 1. Particularly notable is that, while all channels under-predict errors by ~20\%,channel 7 does not perform significantly worse on this metric. Despite the larger RMSE, predicted errors are still reliable. Using a threshold predicted error, channel 7 would be just as usable as others but would just have more predictions discarded.

Figure \ref{r2}b does not, however, indicate if error predictions are reliable across a range of actual accuracies. The RMSE results are consistent with a well-calibrated prediction but provide no information about the resolution of the probabilistic prediction. For this, predictions must be binned by predicted error and then the actual errors for those samples calculated.
 
Figure \ref{r3} provides granular information about the resolution of predictions. For all channels, the idealized perfectly calibrated curve is either within or close to the 95\% confidence bars for smaller predicted errors. Larger errors tend to be systematically under-predicted for all channels, particularly channel 7. However, the point where the experimental results and idealized curve diverge is well into the extremes of the overall error distribution, as is evident from comparing this point to the CDF in blue. Errors are reliably predicted across a range that includes the vast majority of samples. Even once the experimental results do diverge from the idealized curve, the trend is monotonic with larger predicted error standard deviations tending to have larger true errors. 

Benchmarking computational efficiency against CRTM indicates that the ML emulator is significantly faster than CRTM, with this advantage likely to be even larger for cloudy conditions in which CRTM is slower. Comparing the Jacobians of the physics-based CRTM against the emulator for water vapor channels, mean and spread both agree well in the troposphere. Above 200 hPA where there is very little water vapor, the ML emulator has no variability on which to train and therefore is relatively unconstrained. As a result, the Jacobians diverge. However, SHAP analysis of the ML emulator confirms that there is minimal impact from water vapor at these altitudes

\section{Conclusion}
The goal of this work was to train a fast NN emulator of CRTM that could in the future be used to assimilate ABI observations that are currently unused for generating forecast analyses. The NN we have trained is faster (roughly an order of magnitude faster than CRTM based on initial tests), approximately as accurate as other NN CRTM emulators that have been built, and generates reliable probabilistic error predictions. The latter is key for the next steps in moving from research to an operational setting. A threshold can be applied and the NN is exclusively used where the predicted error is low.

The error predictions become less reliable for larger (less frequent) errors. These represent a small fraction of the total number of samples used in the test dataset, and can likely be mitigated if needed with a larger training set. However, given the relative infrequency and the fact that the calibration curves remain monotonic, there is no need to do so before using the threshold method described in the paragraph above. Additionally, in an operational setting standard quality control procedures will likely discard results with excessively large deviations from observations.

The XAI results further support this conclusion. Multiple lines of evidence in the SHAP results indicate that the NN has internalized relevant physics, not just memorized the input dataset. The impact of water vapor at different pressure levels is as expected in the water vapor channels. Ozone mostly impacts the ozone channel. Lastly, channel 7, the only one for which reflected solar IR is relevant, is highly and nearly exclusively impacted by solar angle. All of these provide evidence that the NN is generating its predictions based on physically reasonable patterns and can be relied on to produce physically reasonable predictions on out-of-sample data presented to it in the future.

%

%

\clearpage
\acknowledgments
Support for this work was provided by the Office of Naval Research (ONR) award number N000142312092. This work utilized the Alpine high performance computing resource at the University of Colorado Boulder. Alpine is jointly funded by the University of Colorado Boulder, the University of Colorado Anschutz, Colorado State University, and the National Science Foundation (award 2201538). This work additionally utilized the Blanca condo computing resource at the University of Colorado Boulder. Blanca is jointly funded by computing users and the University of Colorado Boulder. The authors also sincerely thank the anonymous reviewers who provided feedback on the submitted manuscript.


%
%
\datastatement
The test dataset and code used for producing the results and plots in this manuscript are publicly available \citep{howard_probabilistic_2024}.


%






%



 \bibliographystyle{ametsocV6}
 \bibliography{references}

\end{document}